\documentclass[aps,epsfig,manuscript,showpacs]{revtex4}

\usepackage{amsmath,amssymb,graphicx}
\usepackage{epsfig}
\usepackage{graphicx}
\usepackage{enumerate}
\newcommand{\beq}{\begin{equation}}
\newcommand{\eeq}{\end{equation}}

\newcommand{\bea}{\begin{eqnarray}}
\newcommand{\eea}{\end{eqnarray}}

\begin{document}

\begin{large}
{\bf \noindent
Optical conductivity and the correlation strength of high temperature copper-oxide superconductors}
\end{large}

\vspace{0.2in}

\centerline{Armin Comanac$^{(1)}$, Luca de' Medici$^{(2)}$, Massimo Capone$^{(3)}$, A. J. Millis$^{(1)}$ }

\vspace{0.2in}
$^{(1)}$Department of Physics, Columbia University, 538 W.
$120^{th}$ Street, New York, NY 10027 USA

\vspace{0.1in}

$^{(2)}$Department of Physics and Center for Materials Theory, Rutgers the State University of NJ,
136 Frelinghuysen Road, Piscataway, NJ 08854

\vspace{0.1in}
$^{(3)}$SMC, CNR-INFM, and Dipartimento di Fisica, Universitˆ di Roma "La Sapienza", Piazzale A. Moro 2, I-00185, Rome, Italy and ISC-CNR, Via dei Taurini 18, I-00185, Rome, Italy
\date{\today}

%

\vspace{1.5in}
{\bf High temperature copper-oxide-based
superconductivity is obtained by adding carriers
to insulating "parent compounds". It is widely believed the parent compounds
are "Mott" insulators, in which the lack of conduction arises from anomalously
strong electron-electron repulsion, and that the unusual properties of Mott insulators
are responsible for high temperature superconductivity. This paper presents
a comparison of optical conductivity measurements
and theoretical calculations which challenges this belief.
The analysis indicates that  the correlation strength
in the cuprates is not as strong as previously believed,  that the materials are
not properly regarded as Mott insulators, that antiferromagnetism is essential
to obtain the insulating state and, by implication, that antiferromagnetism
is essential to the properties of the doped metallic
and superconducting state as well. }

Since their discovery in 1986, the  high temperature copper-oxide superconductors have
been a central object of study in condensed matter physics. Their highly unusual properties
are widely (although not universally) believed to be a consequence of 
electron-electron interactions which are so strong
that the traditional paradigms of condensed matter physics do not apply: instead, entirely new concepts
and techniques are required to describe the physics. In particular, the high-$T_c$ materials are obtained by
adding carriers to insulating parent compounds such as $La_2CuO_4$. The lattice structure and electron counting
of $La_2CuO_4$ is such that there is an odd number of electrons per formula unit.  Thus,
in the absence of further symmetry breaking,  conventional band theory
would predict that the material is a good metal. $La_2CuO_4$ is however not metallic; it is an insulator
with a gap determined by optical spectroscopy to be approximately $1.8eV$ \cite{Basov05}.

From one perspective the insulating behavior is not surprising. At temperature  $T=0$, $La_2CuO_4$ 
has two-sublattice N\'eel order, so that the {\it magnetic} unit cell contains two formula units and 
thus an even number of electrons, compatible with the  observed insulating behavior.  However, the general
consensus has been that the antiferromagnetic order is  irrelevant. Instead, the 
materials have been identified \cite{Anderson87,Lee05}
as 'Mott insulators": materials in which the electron-electron repulsion is 
so strong that the presence of an electron in one unit cell prevents another electron from entering
that cell, independent of any electronic order. 
(While the cuprates are properly regarded
as "charge-transfer" and not "Mott" insulators in the sense of Ref \cite{SOMCTI}, this issue is not relevant here: the high
energy scale physics and chemistry of transition metal ($Cu$) and ligand ($O$) ions 
produces one band of electrons, with an effective
interaction strength which we aim to  determine.  In particular, optical data show that the nearest
bands (arising in main from the non-bonding oxygen orbitals) are $5-6$ eV removed
in energy, with only a weak absorption tail extending down to the energies of relevance here.  
The issue is discussed in more detail
in the supporting on-line material.)
In a Mott or charge-transfer insulator, a density of one electron per unit cell implies a "jammed" situation:
no electron can move without creating an energetically expensive doubly occupied site.
Removing or adding  electrons creates "holes" or doubly occupied sites, whose motion is not
blocked by the jamming effect but is strongly affected by the nontrivial Mott insulating
background in which it moves \cite{Lee05}.

This paper argues that the experimental evidence is not in agreement with the strong correlation,
"Mott" picture: rather, an intermediate coupling picture is appropriate, in which the antiferromagnetic
order (or correlations) are crucial  to the insulating behavior and, by implication, to the physics of
the doped, superconducting compounds. The important experimental evidence leading
to this conclusion is the optical (frequency-dependent)  conductivity,  $\sigma(\omega)$; the linear response function
connecting a frequency dependent, transverse electric field ${\bf E}$ to the  current ${\bf j}$ it induces. At frequencies
less than the interband threshold the measured conductivity is dominated by processes in which an electron moves from one
unit cell to another. In a Mott insulator, such conductivity processes are  
suppressed by the blocking effect of on-site repulsion \cite{Millis04a}, so that the expected low
frequency spectral weight (integrated optical absorbtion strength) is small. We show
here  that in the high-$T_c$ materials the measured low energy spectral weight is too large to be 
compatible with the Mott (blocking) interpretation of the physics of the cuprates. 

The electronic structure of the cuprates is such that one band (per $CuO_2$ unit) 
crosses the chemical potential; all other bands
are full or empty and may to first approximation be neglected. Electrons moving in the relevant
band are subject to an interaction whose most important component is a repulsion disfavoring
configurations in which two electrons occupy the same site at the same time. This physics may be 
expressed mathematically via the "Hubbard" model of a band of electrons
subject to local correlations. Although the
Hubbard model is  not a fully accurate description of the physics of high temperature superconductors,
it contains the essence of the blocking effect and generally accepted \cite{Lee05} as the basic picture on 
which a more refined description  should be based.
We write the model in a mixed momentum $(k)$ space
position $(i)$ space representation as
\begin{equation}
H=\sum_{k,\sigma}\varepsilon_kc^\dagger_{k,\sigma}c_{k\sigma}+U\sum_i{\hat n}_{i,\uparrow}{\hat n}_{i,\downarrow}
\label{HHub}
\end{equation}
Here ${\hat n}_{i,\sigma}$ is the density operator for electrons of spin $\sigma$
on site $i$  and $\varepsilon_k$ is the dispersion given by local density band calculations.
Small variations among different calculations exist, but all agree within a few 
percent on the values of the parameters important for this study, which are the 
bandwidth $W\approx 3eV$ and the "kinetic energy" $K \approx 0.4eV$. 
For definiteness in this paper we use the $\varepsilon_k$ derived 
from the "downfolding" parametrization of  Ref. \cite{Andersen95}.).

At a density of one electron per cell the ground state of $H$, Eq \ref{HHub} is believed to
be a paramagnetic metal at small $U$ (roughly $U<1eV$) and an antiferromagnetic insulator
at larger U, with a small range of antiferromagnetic metal in between. The key question
is whether the antiferromagnetic order is essential to the insulating nature of the ground state.  To determine this
we turn to the single-site dynamical mean field approximation \cite{Georges96}.
In this approximation, spatial correlations between fluctuations are neglected but temporal fluctuations
on a given site are included exactly. If long ranged antiferromagnetic order is not included in the calculation,
one finds at carrier concentration $n=1$ and temperature $T=0$
a critical value $U_{c2}\approx 1.45 W  $ separating a small $U$ metallic 
phase from a large $U$ insulating phase.  The band theory estimate
$W\approx 3eV$ implies $U_{c2} \approx 4.4eV$. 
This large $U$ phase is identified as a Mott insulator because
an energy gap exists at the chemical potential in the absence of any intersite magnetic correlations.   

We calculated the optical conductivity implied by Eq.  \ref{HHub}, representing
the electric field via  a vector
potential  ${\bf A}$, using the minimal coupling $k\rightarrow k-A$ and standard linear
response theory and multiplying the calculated result (a dimensionless conductance per
$CuO_2$ plane) by the conductance quantum $e^2/\hbar$ and dividing
by the mean $LSCO$ interplane distance $d=6\AA$.  The two  main panels of  Fig 1, which plot the calculated
conductivity for several  carrier concentrations at a value  of $U$  slightly greater
than $U_{c2}$ and one somewhat less.  Consider the $x=0$ results, representative of the 
parent compounds of the high-$T_c$ materials.
The $U>U_{c2}$  calculation reveals Mott insulating behavior: even if magnetic order is neglected
the result is insulating (gap in the conductivity spectrum). Adding antiferromagnetism increases the
gap and produces structure at the gap edge. On the other hand, the $U<U_{c2}$ calculation
reveals metallic behavior (no gap) in the absence of antiferromagnetism while the antiferromagnetic
calculation reveals a large gap.

To interpret the results we note that in  models such as 
Eq \ref{HHub} the optical conductivity
obeys a "restricted f-sum rule" \cite{Maldague77,Millis04a}.  Defining 

\begin{equation}
K(\Omega)=\left(\frac{V_{cell}}{a^2}\right)\int_0^\Omega \frac{2d\omega}{\pi}\frac{\sigma(\omega)}{\sigma_Q}
\label{partialsum}
\end{equation}
we have  
\begin{equation}
K(\infty)=\sum_{k,\sigma}n_{k,\sigma}\frac{\partial^2\varepsilon_k}{\partial k_x^2 }
\label{kdef}
\end{equation}
Here $\sigma_Q=e^2/\hbar$ is the conductance quantum,  
$V_{cell}$ is the volume of the unit cell, $a$ is the in-plane lattice constant and $n_{k,\sigma}$ is the probability  that 
the state of momentum $k$ and spin $\sigma$ is occupied.  Note that $\sigma(\omega)$ in Eq \ref{partialsum}
refers to the real (dissipative) conductivity calculated from Eq \ref{HHub}; in physical terms
it corresponds to that contribution
to the measured conductivity arising from transitions within the band of states described by Eq \ref{HHub}. In a real
material, interband transitions not described by Eq \ref{HHub} 
also contribute to the conductivity; these make up the  difference between Eq \ref{kdef}
and the familiar f-sum rule $\int_0^\infty d\omega \sigma(\omega)=\pi ne^2/2m$.

If $U=0$, $n(k,\sigma)$ is  the usual Fermi-Dirac distribution, corresponding to filling
up only the lowest-lying states in the band. Evaluation of Eq \ref{kdef} for this case  yields
$K=K_{band}\approx0.4eV$, essentially independent of carrier concentration for the dopings relevant
to high temperature superconductivity.  In this noninteracting case the electrons are not scattered;
they are freely accelerated by an applied electric field so the conductivity is just a delta
function of strength $\pi K$ at $\omega=0$. Increasing the interaction causes electron-electron
scattering which shifts spectral weight from $\omega=0$ to higher frequencies. Increasing the interaction also  
tends to localize the electrons, leading to an $n(k)$ more uniformly distributed over the band and 
thus reducing the magnitude of the integral  in Eq \ref{kdef}, i.e. decreasing the total spectral weight. 
However, adding holes allows carrier motion, thus increasing the spectral weight and shifting it back towards
$\omega=0$. These effects can be seen in the insets of  Fig. 1 which plot 
the conductivity integral, Eq \ref{partialsum} obtained from the calculated conductivities 
shown in the main panel of the figure.

We now compare the calculation to measurements of the conductivity, of which a representative example \cite{Uchida91}
is shown in Fig. 2. These data were taken in 1991; subsequent improvements especially in sample
quality have sharpened the band gap seen in the conductivity of the $x=0$ sample, so that the onset of absorbtion
begins at $\omega \approx 1.8eV$ but have not changed the material features; in particular the spectral weights
in the different frequency regimes.  Use of the band theory estimate $W\approx 3eV$ would imply the band
gap is approximately $0.6W$, consistent with the result of the antiferromagnetic-phase $U<U_{c2}$ calculation
but inconsistent with the antiferromagnetic phase $U>U_{c2}$ result.  The antiferromagnetic
$U>U_{c2}$ calculation can be made consistent with the observed band gap by reducing the energy scales
by $25\%$, implying in particular a bandwidth  $W^*\approx 2.25eV$ instead of the $W\approx 3eV$
found in band theory calculations.  However, even  if this renormalization is made,  the
magnitude of the observed conductivity is inconsistent with the $U>U_{c2}$ hypothesis, as will now be shown.

The measured spectral weight in the range 
$\omega<3eV$  for the insulating compound corresponds to $K(3eV)=0.2eV$ or about $50\%$ of the noninteracting value.
It is likely that not all of the spectral weight observed in the range below $3eV$ is due
to the optical transitions of interest. Interband transitions to irrelevant bands may contribute.   
To obtain an upper bound on possible  interband contributions
we note that as doping increases the calculated conductivity shifts strongly to lower frequencies
(as may be seen in Fig. 1). We therefore
use the measured $x=0.34$ data in the range $\omega>1eV$ as an estimate of the interband 
contribution to the conductivity. We have integrated the difference between the conductivity measured
in the $x=0$ sample and that measured in the $x=0.34$ sample over the range $\omega<3eV$
obtaining $K_{exp}(\Omega=3eV) \approx 0.2K_{band}\approx 0.1eV$. This estimate is quite consistent
with the results shown in the inset of the lower panel of Fig 1. However, the $U>U_{c2}$ calculation
yields substantially less spectral weight in the low frequency regime.  Combining the band theory
estimate $W\approx 3eV$  with the data in the inset of the upper panel of Fig. 1 yields  $K(3eV)\approx 0.03eV$, 
far less than the measured $0.1eV$. If we use instead the renormalized $W^*=2.25eV$ which reproduces the value of the gap,
then $3eV \approx 4W^*/3$. The total spectral weight
integrated up to this point is $0.13K(U=0)$ and because the theoretical $K(U=0) \sim W$ this implies
an integrated weight of about $0.05eV$ still much smaller than what is observed.

Now $La_2CuO_4$ is observed to remain 
insulating at temperatures above its N\'eel temperature $\approx 340K$, so {\em long ranged} order
is not essential to the insulating behavior. However, the Ne\'el temperature is strongly suppressed by low
dimensional fluctuation effects and is a poor measure of the strength of the magnetic correlations,
which are found to remain significant up to the highest measured temperatures ($T \approx 1000K$)
\cite{Imai93}. Recent cluster dynamical 
mean field calculations(K. Haule et. al., private communication; E. Gull, P. Werner, M. Troyer and A. Millis, 
to be published ) produce insulating
behavior over wide temperature ranges  without long ranged order, even in the intermediate coupling regime,
provided that near-neighbor spin correlations are strong enough.

We next turn to the doping dependence of the conductivity. The solid symbols in Fig. 3. show 
the optical spectral weight for several cuprate
materials, integrated up to $0.8eV$, about $0.45$ of the insulating gap.  The value is chosen
because available evidence indicates that the conductivity at $\omega<1eV$ is essentially
uncontaminated by interband transitions while at higher frequencies the situation is less clear \cite{Millis05}.
One sees that the measured spectral weight
in the mid-gap region scales linearly with doping, but with a non-vanishing intercept. The open symbols show
the results of the theoretical calculations for a $U$ slightly greater than $U_{c2}$ and for two
$U$-values less than $U_{c2}$. For the $U>U_{c2}$ calculation we have used the scale $W^*=2.25eV$
to convert the theoretical results to physical units. 
We see that for $U>U_{c2}$ the calculated spectral weight  is qualitatively
inconsistent with the data, because it vanishes as doping tends to zero. 
However, we  note that in the qualitative comparison  
the decisive feature is the behavior at $x<0.1$ 
where the uncertainties in the data are largest.
Further experimental examination of this frequency regime would  be desirable. 
The $U>U_{c2}$ results are  also somewhat smaller in magnitude
than the experimentally determined values.  On the other hand, the results
for $U=0.9U_{c2}$   give a magnitude and doping dependence which is reasonably
consistent with the measured values at non-vanishing dopings. 
The $x>0$ calculations are performed within single-site dynamical mean field theory in the paramagnetic
phase. This  method does not take into account the effects of near-neighbor magnetic correlations,
which are likely to be present even in the absence of true long ranged order and which will suppress somewhat
the spectral weight in the low frequency regime. We suggest that a model with a $U \approx 0.85U_{c2}$
and with a proper treatement of antiferromagnetic correlations will lead to a doping dependence
of the spectral weight which is consistent with experiment.  

The results presented here suggest that a re-examination of theoretical approaches to high temperature
superconductivity would be worthwhile. Much work has been based on the "t-J'' model \cite{Lee05},
which is derived on the assumption that the correlation-induced blocking effect is fundamental, with 
antiferromagnetism providing a next correction and which has been widely accepted because it provides
natural explanation of the  striking doping dependence of physical properties. Determining whether the 
observed doping dependence of the low energy physics can be understood within the intermediate-coupling,
strong antiferromagnetic correlations picture implied by the optical data, is an important challenge for future work.

{\it Acknowledgements} This collaboration was begun with support from the 
Columbia/Polytechnique/Science Po /Sorbonne Alliance 
program.  AJM and AC were supported by  NSF-DMR-0705847 and
M.C.  by MIUR PRIN 2005, Prot. 200522492. {\it Correspondence} should be addressed to 
A. J. Millis (millis@phys.columbia.edu)

\newpage

{\bf Figure Captions}

{\bf Fig. 1: Optical conductivity of Hubbard model:} calculated as described in main text at dopings $x=1-n$
and interaction strength $U$ indicated. Main panels:
optical conductivity; insets; optical integral. For $x=0$ both paramagnetic and antiferromagnetic
phase calculations are shown; for $x>0$ only paramagnetic phase results are given. If the band theory
value $W=3eV$ is used then the frequency scale is electron volts.

{\bf Fig. 2: Measured optical conductivity of $La_{2-x}Cu_xCuO_4$} reproduced from
Ref. \cite{Uchida91}); solid vertical line at frequency $\omega=0.8eV$ indicates
cutoff frequencies used for spectral weight analysis.

{\bf Fig. 3: Comparison of measured and calculated optical spectral weight}.  
Solid symbols: spectral weight obtained by integrating experimental conductivity
up to $0.8eV$ from references given. Open symbols: theoretically calculated spectral weight, integrated up to $W/4$. 
For $U=0.85U_{c2}$ and $U=0.9U_{c2}$
the band-theory estimate $W=3eV$ is used to convert the calculation to physical units; for
$U=1.02U_{c2}$ the value $W=2.25eV$ which reproduces the insulating gap is used.


\newpage
{\bf \large Fig. 1}

\vspace{.25in}

\begin{figure}[htb]
\includegraphics[width=0.75\linewidth]{u6inclafm.eps}

\vspace{.5in}
\includegraphics[width=0.75\linewidth]{u5inclafm.eps}
\label{fig:u6t}
\end{figure}

\newpage
{\bf \large Fig. 2}

\vspace{1in}

\begin{figure}[htb]
\includegraphics[width=\linewidth]{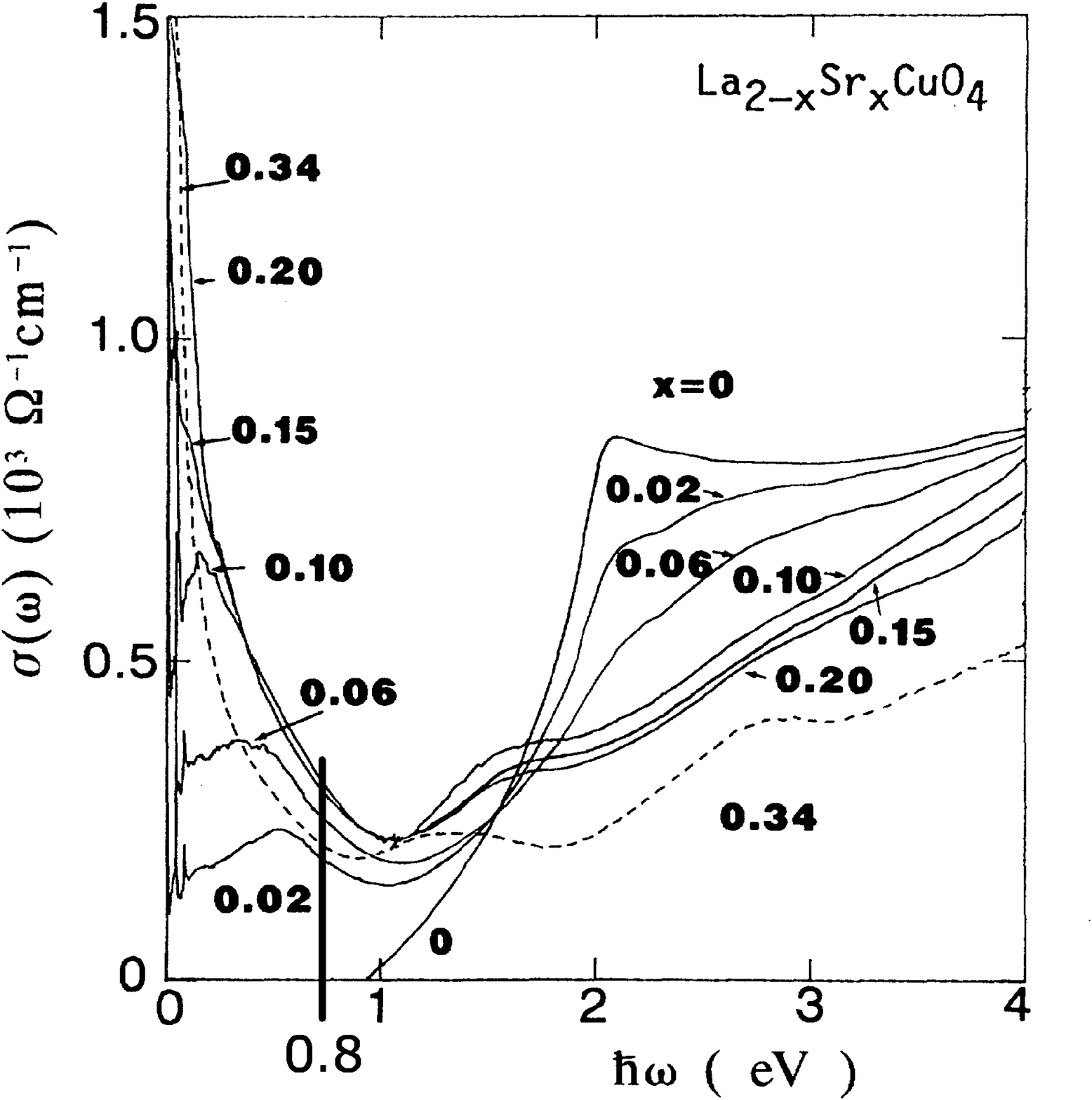}
\label{fig:exp_cond}
\end{figure}

\newpage

{\bf \large Fig. 3}

\vspace{1in}

\begin{figure}[htb]
\includegraphics[width=\linewidth]{CompExp11.eps}
\label{fig:comp_exp}
\end{figure}

\end{document}